# A-BASE-DE-PROS: una implementación práctica de los Objetivos de Desarrollo Sostenible en la Universidad Politécnica de Madrid

Patricia Almendros[1] -p.almendros@upm.es-, Silvia Otegui[1] -s.otegui@alumnos.upm.es-, Alejandro Nares[1] -alejandro.nares.sanchez@alumnos.upm.es-, Laura del Fresno[1] -laura.delfresno@upm.es-, Javier Ablanque[1] -j.ablanque@alumnos.upm.es-, Irene Blanco[1] -irene.blanco@upm.es-, Juan Ramón Ferrer[1] -juanramon.ferrer@upm.es-, Sonia Benito[1] -sonia.benito@upm.es-, Carmen Lopez[1] -carmen.lopez@upm.es-, Sonia García[1] -sonia.garcia@upm.es-, León Fernández[1] -leon.fernandez@upm.es-, Sergio Zubelzu[1] -sergio.zubelzu@upm.es-, Raúl Sánchez[1] -raul.sanchez@upm.es -, Paloma Esteve[1] -paloma.esteve@upm.es-, Rosa María Benito[1] -rosamaria.benito@upm.es-, Juan Carlos Losada[1] -juancarlos.losada@upm.es-, Antonio Saa[1] –antonio.saa@upm.es-, Gabriel Gascó[1] -gabriel.gascó@upm.es-, Ana M Méndez[2] -anamaria.mendez@upm.es-, Mónica Montoya[1] -monica.montoya@upm.es-, Marina de Francisco[1] -marina.defrancisco@upm.es-, Jesús Ruiz[3] -jesus.ruiz@elporvenir.es-, Samuel Seoanez[3] -samuel.seoanez@elporvenir.es-, Sara Castilla[3] -sara.castilla@elporvenir.es-, Dámaris Fuente[3] -damaris.fuente@elporvenir.es-, Raquel Santiuste[3] -raquel.santiuste@elpovenir.es-, María Dolores Polo[3] -lola.polo@elporvenir.es-, Juan Carlos Viana[3] -jcarlos.viana@elporvenir.es-, Alberto Serrano[4] -albertosh050996@gmail.com-, Raimundo García[5] -raimundo.garcia@maristaschamberi.com-, Sergio Burgas Fernández[5] -sergio.burgas@maristaschamberi.com-, Aurora Gutiérrez[5] -aurora.gutierrez@maristaschamberi.com-, Pablo Pascual[5] -pablo.pascual@maristaschamberi.com-, Beatriz Parra[5] -beatriz.parra@maristaschamberi.com-, Manuel Soto[6] -manuelsotogarde@gmail.com-, María Felisa Domínguez[6] -mdr.educamadrid@gmail.com-, Cristina Martín[6] -cristinamartin319@hotmail.com-, Cristina López[6] -cristinamcprevencion@gmail.com-, Juan Carlos Alarcón[6] -jcaees@gmail.com-, Nuria Alfonso Lidón[7] -nuria.alfonsolidon@educa.madrid.org-, Esther Garde[8] -esther.garde@educa.madrid.org-, María José Maroto[7] -mariajose.marotovidal@educa.madrid.org-, Elena Esteban[9] -elenaestebanandres@gmail.com-, Lucía Veguillas[9] -luciasanjaimeapostol@gmail.com-, Francisco Mora[9] -colegiosanjaimeapostol@gmail.com-, María del Prado Montalvo[10] -mmontalvo@iesjulioverne.org-, Olga Villanueva[10] -ovillanueva@iesjulioverne.org-, Elena Villa[10] -evilla@iesjulioverne.org-, Fabio Revuelta[1] -fabio.revuelta@upm.es-

[1] Escuela Técnica Superior de Ingeniería Agronómica, Alimentaria y de Biosistemas, Universidad Politécnica de Madrid. Avenida Puerta de Hierro 2-4, 28040 Madrid.

[2] Escuela Técnica Superior de Ingenieros de Minas y Energía, Universidad Politécnica de Madrid. Calle Ríos Rosas 21, 28003 Madrid.

[3] Colegio El Porvenir. Calle Bravo Murillo 85E, 28003 Madrid.

[4] Instituto de Enseñanza Secundaria Pradolongo, Calle Albardín 6, 28026 Madrid.

5 Colegio Maristas Chamberí, Calle de Rafael Calvo 12, 28010 Madrid.

[6] Instituto de Enseñanza Secundaria Rafael Frühbeck, Calle de Humanes 12, 28914 Leganés.

[7] Instituto de Enseñanza Secundaria San Nicasio, Paseo de la Hermita, 28914 Leganés.

[8] Instituto de Enseñanza Secundaria Altaír, Carretera de Getafe a Leganés s/n, 28904 Getafe.

9 Colegio San Jaime Apóstol, Calle de Juan José Martínez Seco 54, 28021 Madrid.

10 Instituto de Enseñanza Secundaria Julio Verne, Calle Ingeniería, 4, 28918 Leganés.

## RESUMEN

A lo largo de los últimos años hemos visto cómo los Objetivos de Desarrollo Sostenible (ODS) han ido permeando las distintas capas de la sociedad, estableciendo nuevas prioridades tanto en las políticas públicas como en las empresariales. La educación no ha quedado ajena a este cambio, sino que también se está alineando con las metas anteriores. En este capítulo se describen las principales actividades realizadas en el marco del proyecto APS22.2003 "Aprendizaje BAsado en SErvicio DE ODS relacionados con una PROducción y consumo responsableS (A-BASE-DE-PROS)", en el que el ODS 12 se toma como eje central para concienciar a estudiantes universitarios y de secundaria de la importancia de la Agenda 2030. En general, hemos comprobado que el proyecto ha permitido dar a conocer los ODS y que la mayoría de los estudiantes, tanto universitarios como de secundaria, han valorado la experiencia como positiva.

## Palabras clave

Objetivos de Desarrollo Sostenible, consumo responsable, educación superior, educación secundaria

## ABSTRACT

The influence of the Sustainable Development Goals (SDGs) has been widely spread over the last years, establishing new public and privat policies. Education has also been experiencing this change by aligning with the previous goals. In this chapter, we briefly summarize the main activities conducted under the Grant APS22.2003 "Service-based learning of the SDGs related to a responsible production and consumption (A-BASE-DE-PROS)", which uses the SDG 12 as a guide line to raise the awareness of the importance of the 2030 Agenda among undergraduate and secondary-school students. In general, the service-based learning has increased the knowledge of the SDGs among the students. Furthermore, most of the (university and secondary) students found the service-learning experience positive.

## Keywords

Sustainable Development Goals, responsible consumption, high education, secondary education

## 1. INTRODUCCIÓN

Los Objetivos de Desarrollo Sostenible (ODS) son 17 ambiciosas metas que forman parte de la Agenda 2030 que la Organización Mundial de las Naciones Unidas (ONU) estableció el 25 de septiembre de 2015 para hacer de nuestro planeta un lugar más justo, seguro y sostenible (ODS, 2015) (ver Figura 1). Su influencia en nuestra sociedad actual resulta innegable y podemos ver referencias a los ODS prácticamente cada día en las campañas publicitarias de numerosas empresas, actos reivindicativos de organizaciones y colectivos sociales e incluso en las acciones de los gobiernos públicos, tanto a nivel regional como nacional e internacional. Aunque los ODS tienen un carácter muy amplio (como erradicar la pobreza en el caso del ODS 1 o conseguir comunidades y ciudades sostenibles en el caso del ODS 11), contienen también una serie de descriptores específicos que permiten cuantificar en qué medida los distintos países están alcanzando las diferentes metas.

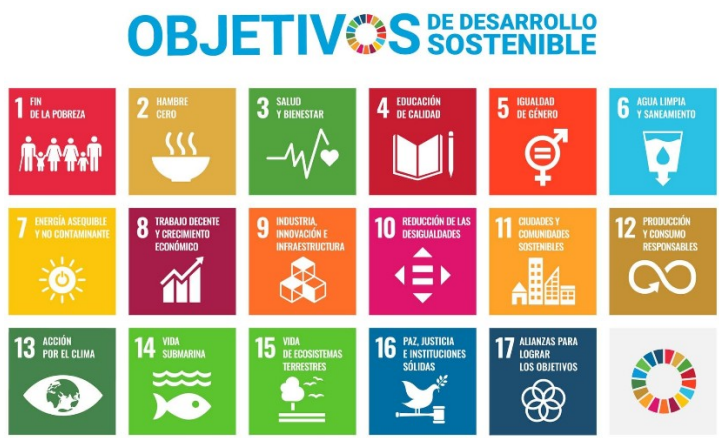

Figura 1. Los 17 Objetivos de Desarrollo Sostenible establecidos en el año 2015 por la Organización de Naciones Unidas (ODS, 2015).

Aunque la ONU estableció un plazo de 15 años para alcanzar completamente todos los ODS, se espera que éstos tengan un claro carácter a más largo plazo (SDR, 2021). En consecuencia, resulta imprescindible que los jóvenes los conozcan y tomen conciencia de su importancia, dado que serán ellos quienes más se beneficien de su consecución (o quienes más sufran las consecuencias de no alcanzarlos). Por ello, los centros educativos resultan un entorno óptimo para dar a conocer los ODS, permitiendo, además, implementar en el aula las relaciones ciencia-tecnología-sociedad con una perspectiva *STEM (Science-Technology-Environment-Mathematics)* interdisciplinar (Widya, 2019), lo que suele aumentar la motivación e implicación del alumnado al permitirle ver cómo se aplica lo estudiado en el aula en su vida diaria. De hecho, la última reforma de la educación secundaria incorpora los ODS en el perfil de salida de los estudiantes de 4º de Educación Secundaria Obligatoria (ESO), apareciendo, además, en dos de las competencias (competencia ciudadana y competencia digital) que los egresados deben haber adquirido, tanto en la ESO (RD 217/2022) como en el Bachillerato (RD 243/2022). En el caso de la educación superior, no hay una normativa tan clara que se refiera a la implementación de los ODS, pero el Ministerio de Transición Ecológica y el Reto Demográfico ha elaborado una guía con consejos sobre cómo incorporar los ODS en la universidad (Kestin, 2017). Además, numerosas instituciones han creado planes de sostenibilidad y portales para dar a conocer a la sociedad las acciones que están llevando a cabo para trabajar los ODS, como el Portal de Sostenibilidad de la Universidad Politécnica de Madrid (UPM) (UPM, 2023).

El aprendizaje basado en servicio (ApS) ofrece un marco único para desarrollar los ODS entre el alumnado. Esta metodología ha ido adquiriendo cada vez más importancia para aumentar la motivación e implicación de los estudiantes, tanto en educación secundaria como en educación superior. En el ApS, el estudiante lleva a cabo su *aprendizaje* típicamente fuera del aula realizando alguna actividad de *servicio* real en su entorno (Resch, 2021). Así, se han llevado numerosas experiencias en las que los estudiantes aprendían reparando los patios de sus colegios, colaborando con organizaciones no gubernamentales o bien rehabilitando parques de cercanos, por citar tan sólo tres ejemplos.

La eficacia del ApS radica en dos hechos. Por un lado, la puesta en práctica del conocimiento y las destrezas adquiridas en el aula sirven para fijar los conceptos y destrezas adquiridos, al llevarse a cabo por medio de un aprendizaje activo que es, en general, siempre más significativo, profundo y duradero. Por otro lado, el ApS favorece el desarrollo de competencias vinculadas a cuestiones sociales y éticas, así como el desarrollo de numerosas competencias transversales que van desde el trabajo colaborativo y comunicación oral y escrita al liderazgo y la gestión de equipos, entre otros. Por ello, el proyecto APS22.2003 "Aprendizaje BAsado en SErvicio DE ODS relacionados con una PROducción y consumo responsableS (A-BASE-DE-PROS)" ha tomado como eje central el ODS 12 de producción y consumos responsables para concienciar a estudiantes, tanto universitarios como preuniversitarios, de la importancia de la conservación del medio ambiente (ver Figura 2). Se ha seleccionado el ODS 12 puesto que es el que los estudiantes pueden trabajar de una forma más clara en su día a día. No obstante, este ODS no forma un ente aislado sino está fuertemente relacionado con otros como el ODS 6 referente al uso y distribución de agua, el ODS 7 relacionado con el consumo y generación de energía o los ODS 13 y 14 referidos al impacto sobre la naturaleza de las actuaciones del ser humano (lo que enlaza, también, con el ODS 8 de trabajo decente).

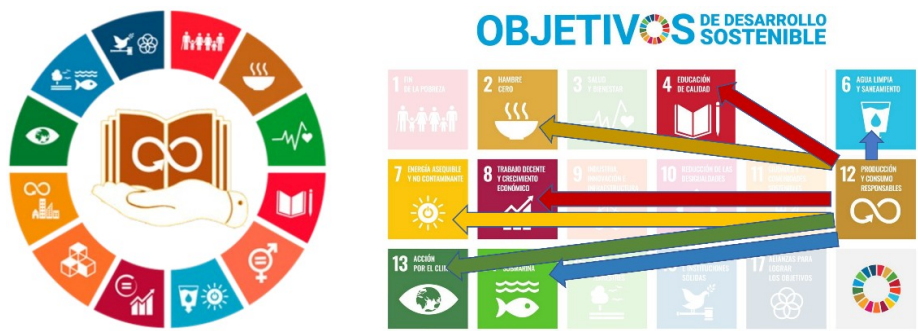

Figura 2. El proyecto APS22.2003 "Aprendizaje BAsado en SErvicio DE ODS relacionados con una PROducción y consumo responsableS (A-BASE-DE-PROS)", cuyo logo se muestra a la izquierda, desarrolla el Objetivo de Desarrollo Sostenible (ODS 12) de producción y consumo responsables (derecha) a la vez que conciencia a los estudiantes de la importancia del consumo razonable de agua (ODS 6) y energía (ODS 7) por su influencia en el medio ambiente (ODS 13 y ODS 14), además de hambre (ODS 2) y del trabajo y el crecimiento económico (ODS 8), tratando, de forma implícita, el ODS 4 de educación de calidad.

En este capítulo se resumen brevemente las principales características del proyecto A-BASE-DE-PROS. El capítulo está organizado de la siguiente manera. A continuación, se describe el caso de estudio en el que se ha llevado a cabo la experiencia. En la sección 3 presentamos algunas de las actividades que se han llevado a cabo. Por último, en la sección 4 resumimos las principales conclusiones extraídas.

## 2. CASO DE ESTUDIO

El ApS A-BASE-DE-PROS se ha llevado a cabo en la UPM durante el año 2022. La Tabla 1 muestra las asignaturas en las que se ha llevado a cabo el ApS, la mayoría de ellas asociadas a alguno de los 5 grados *verdes* de la ETS Ingeniería Agronómica, Alimentaria y de Biosistemas (ETSIAAB), si bien han participado también una asignatura de la ETS Montes, Forestal y del Medio Natural y dos asignaturas de postgrado. En total, han colaborado 14 profesores de tres centros. Dado el alto número de estudiantes matriculados en algunas de las asignaturas participantes (más de 700 en total, siendo su mayoría estudiantes de grado) únicamente 50 han llevado a cabo todas las actividades asignadas, si bien otros muchos han participado en alguna de las actividades de los ApS. El *servicio* se ha realizado sobre más de 600 estudiantes de educación secundaria de alguno de los 6 centros de la Comunidad de Madrid implicados (ver Figura 3), que eran tanto públicos (Institutos de Enseñanza Secundaria Rafael Frühbeck de Burgos, Julio Verne y San Nicasio, de Leganés) como privados o concertados (Colegios El Porvenir, Maristas Chamberí y San Jaime Apóstol, de Madrid).

Tabla 1. Características de las asignaturas que han participado en el proyecto APS22.2003 "Aprendizaje BAsado en SErvicio DE ODS relacionados con una PROducción y consumo responsableS (A-BASE-DE-PROS)".

|  | Asignatura | Nº Alumnos | Curso | Semestre | Tipo | Centro |
|---|---|---|---|---|---|---|
|  | **GRADO EN INGENIERÍA AGROAMBIENTAL** | | | | | |
| 1 | Gestión, Tratamiento y Utilización de Residuos | 21 | 3 | 6 | OBL | Escuela Técnica Superior de Ingeniería Agronómica, Alimentaria y de Biosistemas (ETSIAAB) |
| 2 | Evaluación, Conservación y Recuperación de Suelos y Aguas | 11 | 4 | 7 | OPT | ETSIAAB |
| 3 | Economía General | 75 | 1 | 2 | OBL | ETSIAAB |
| 4 | Hidráulica e Hidrología Ambiental | 25 | 3 | 6 | OBL | ETSIAAB |
| 5 | Física I | 82 | 1 | 1 | OBL | ETSIAAB |
|  | **GRADO EN INGENIERÍA AGRÍCOLA** | | | | | |
| 6 | Química Aplicada a la Ingeniería Agrícola | 128 | 1 | 2 | OBL | ETSIAAB |
| 7 | Degradación de Agrosistemas y Cambio Climático. | 79 | 3 | 5 | OBL | ETSIAAB |
| 8 | Electrificación de Explotaciones Agropecuarias | 25 | 2 | 6 | OBL | ETSIAAB |
|  | **GRADO EN INGENIERÍA ALIMENTARIA** | | | | | |
| 9 | Gestión y Aprovechamiento de Residuos | 56 | 4 | 8 | OBL | ETSIAAB |
|  | **GRADO EN CIENCIAS AGRARIAS Y BIOECONOMÍA** | | | | | |
| 10 | Administración de Empresas | 38 | 2 | 4 | OBL | ETSIAAB |
|  | **GRADO EN BIOTECNOLOGÍA** | | | | | |
| 11 | Economía y Gestión de Empresas | 97 | 2 | 3 | OBL | ETSIAAB |
|  | **GRADO EN INGENIERÍA EN TECNOLOGÍAS MEDIOAMBIENTALES** | | | | | |
| 12 | Tecnología para el Tratamiento de Suelos Contaminados. | 61 | 3 | 5 | OBL | ETS Montes, Forestal y del Medio Natural |
|  | **MÁSTER EN INGENIERÍA AGRONÓMICA** | | | | | |
| 13 | Reutilización y Valorización de Subproductos | 18 | 2 | 4 | OBL | ETSIAAB |
|  | **MÁSTER UNIVERSITARIO EN FORMACIÓN DEL PROFESORADO DE ESO, BACHILLERATO Y FP** | | | | | |
| 14 | Didáctica en Física y Química | 28 | 1 |  | OBL | ETSIAAB/ICE |

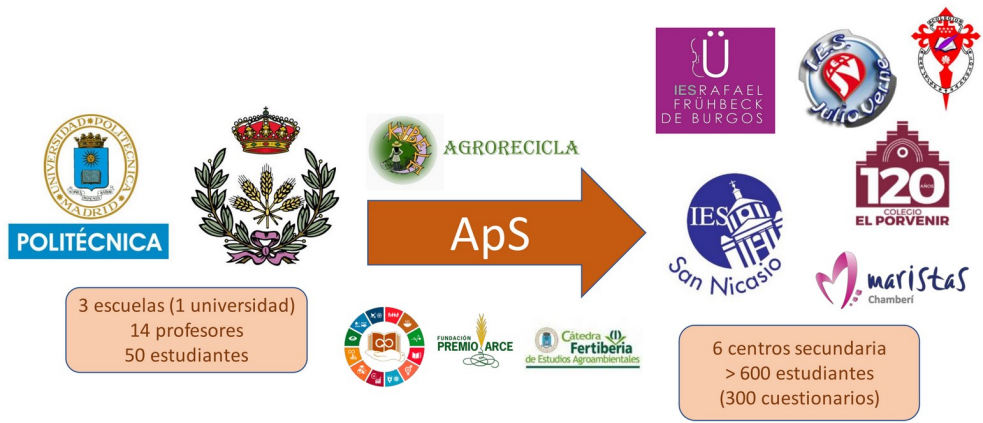

Figura 3. 14 profesores y 50 estudiantes universitarios de tres centros de la Universidad Politécnica de Madrid (fundamentalmente de la Escuela Técnica Superior de Ingeniería Agronómica, Alimentaria y de Biosistemas) han llevado a cabo el aprendizaje basado en servicio con más de 600 estudiantes de secundaria de 6 centros de la Comunidad de Madrid.

## 3. ACTIVIDADES REALIZADAS Y EVALUACIÓN DE LA EXPERIENCIA

En esta sección se presentan algunas de las actividades que los estudiantes UPM han realizado con los estudiantes de secundaria. La Figura 4 muestra un breve resumen de algunas de ellas. Como hemos mencionado anteriormente, aunque el ODS 12 ha sido el hilo conductor de todo el ApS, las actividades que se han llevado a cabo también se relacionan con otros ODS. Por ejemplo, la actividad sobre cultivos hidropónicos ha permitido trabajar otros ODS que están fuertemente conectados con cómo garantizar modalidades de consumo y producción sostenibles (ODS 12), como el ODS 6, agua limpia y saneamiento, o el ODS 2, hambre y seguridad alimentaria.

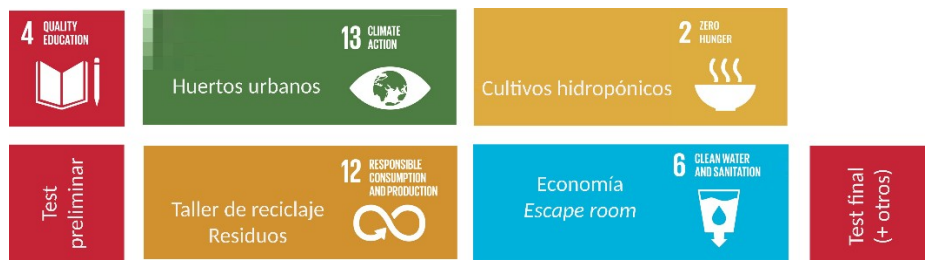

Figura 4. Resumen de algunas de las actividades realizadas para tratar los diferentes Objetivos de Desarrollo Sostenible.

### 3.1 Actividades realizadas

#### 3.1.1 ODS 6. La importancia del agua en nuestras vidas

##### 3.1.1.1 La gran pestilencia y las políticas públicas del agua

Algunos de los estudiantes de las asignaturas Economía General (Grado en Ingeniería Agroambiental) y Administración de Empresas (Grado en Ciencias Agrarias y Bioeconomía) analizaron la sostenibilidad y la economía circular del agua a través del evento de la gran pestilencia *(the great stink),* que tuvo lugar en 1858 en Londres. En ella, la contaminación y el bajo nivel del río Támesis dio lugar a una fuerte crisis sanitaria en la que el cólera y el tifus ocasionaron miles de muertos. Los alumnos trabajaron en grupo analizando las causas, consecuencias y posibles soluciones económicas y ambientales. Los mejores trabajos realizados por los estudiantes fueron seleccionados para participar en el ApS en el Colegio El Porvenir. En este centro, después de presentar la UPM y el ODS 6 a un grupo de 66 estudiantes de secundaria, éstos fueron agrupados en grupos más pequeños a los que los estudiantes universitarios presentaron sus trabajos. Los estudiantes universitarios se encargaban, además, de supervisar una competición entre los estudiantes de secundaria basada en un cuestionario de Kahoot! mostrado en la Figura 5.

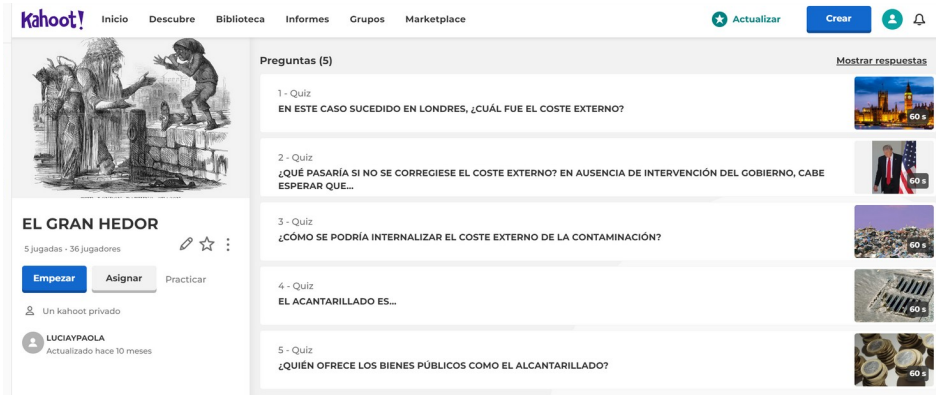

Figura 5. Kahoot! preparado por los estudiantes universitarios y realizado con los estudiantes de secundaria del Colegio El Porvenir.

### 3.1.1.2 Escape room del agua

Se ha diseñado una *escape room* para tratar el ODS 6 de agua limpia y saneamiento (ver Figura 6). En esta actividad, los alumnos de ESO tenían que completar 6 pruebas bajo la supervisión de estudiantes universitarios, que actuaban como monitores. Cada vez que respondían a una de las pruebas en la pantalla se mostraba un mensaje como alguno de los mostrados en la Figura 6. Si lo hacían correctamente, se les daba una pista con la que podían completar las palabras de un crucigrama. Una vez completo este último, los estudiantes de secundaria tenían que resolver el enigma final combinando algunas de las letras que aparecían en las casillas del crucigrama coloreadas en amarillo, verde y azul (solución: ODS + AGUA + SANEAMIENTO).

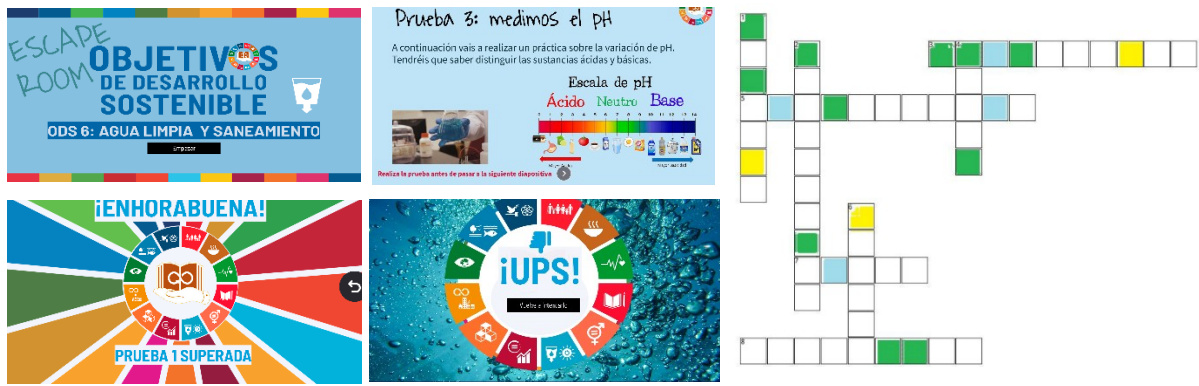

Figura 6. Capturas de pantalla de la presentación usada en la *escape room* diseñada para tratar el ODS 6 de agua limpia y saneamiento (arriba izquierda, portada; arriba centro, diapositiva de una de las pruebas; abajo izquierda, mensaje mostrado al completar correctamente una de las pruebas; abajo centro, mensaje mostrado al contestar mal una de las pruebas) y crucigrama que debían completar para resolver el enigma final (izquierda).

### 3.1.2. ODS 12. El tratamiento de residuos y el consumo responsable

La asignatura de Gestión y Aprovechamiento de Residuos del Grado en Ingeniería Alimentaria tiene como uno de sus objetivos principales fomentar el respeto por el medio ambiente y el uso sostenible de recursos. Para lograr este objetivo, los estudiantes de la misma tuvieron que llevar a cabo una valorización para transformar residuos alimenticios en bioplásticos algo, que, posteriormente, explicaron a los estudiantes de secundaria en el marco del ApS (ver Figura 7).

### 3.1.3. ODS 2. Los cultivos hidropónicos contra el hambre

El cambio climático ha hecho que haya aumentado considerablemente a lo largo de los últimos años el interés por los cultivos hidropónicos debido, entre otras causas, a un mejor aprovechamiento de los recursos hídricos, un aspecto que limita considerablemente la producción agraria en muchos países del mundo, ocasionando hambrunas y desnutrición (ODS 2 hambre cero). Para formar a los estudiantes en este tipo de cultivos, algunos estudiantes universitarios prepararon disoluciones nutritivas que posteriormente utilizaron los estudiantes del IES Julio Verne en sistemas de hidroponía construidos por ellos con materiales reciclados.

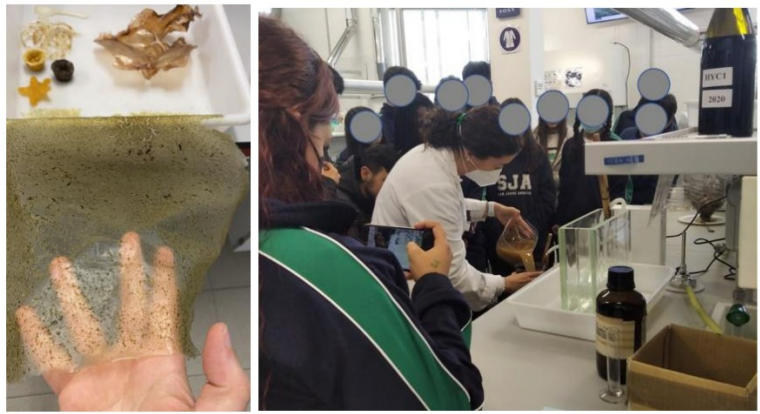

Figura 7. Bioplásticos sintetizados (izquierda) y alumnos de secundaria (derecha) en el laboratorio viendo cómo se trata el agua.

### 3.1.4. ODS 12 y 13. Talleres realizados por asociaciones de estudiantes

Durante las visitas de los estudiantes de secundaria a la ETSIAAB (ver Figura 8), las asociaciones de estudiantes KYBELE y Agrónomos Recicla colaboraron en la visita a los campos de prácticas, el Aula Verde y a los diferentes huertos urbanos con los que cuenta la escuela. Asimismo, los estudiantes de estas asociaciones trabajaron como monitores en dos talleres explicando cómo hacer un huerto urbano (asociación KYBELE) y cómo reciclar correctamente (asociación Agrónomos Recicla).

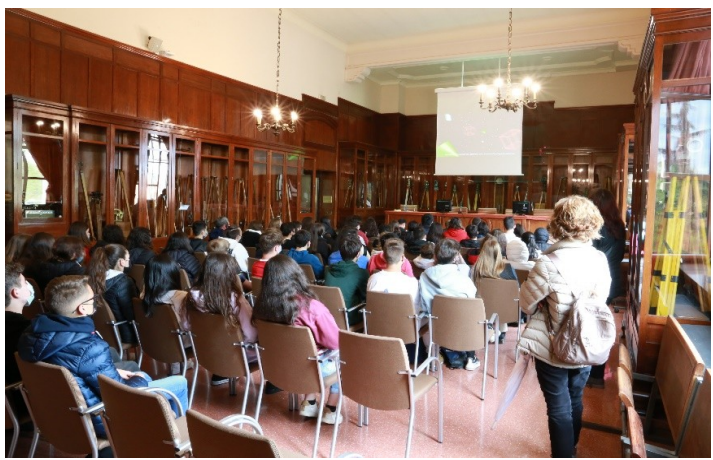

Figura 8. Estudiantes de secundaria durante una visita a la ETS Ingeniería Agronómica, Alimentaria y de Biosistemas.

## 3.2  Evaluación de las actividades

Se ha diseñado un cuestionario para evaluar las actividades realizadas en el marco del ApS que tenían que responder los participantes, tanto los estudiantes universitarios, en calidad de ejecutores del ApS, como los de secundaria, como destinatarios finales del *servicio*. Este cuestionario, que debía cumplimentarse antes y después de llevar a cabo las distintas actividades del ApS, estaba formado por 20 preguntas relacionadas con distintas cuestiones medioambientales implicadas en los ODS, como son qué envases se reciclan en el contenedor amarillo o cuáles son las energías renovables más utilizadas, por citar sólo dos ejemplos. También contenía preguntas para evaluar el posible cambio de hábitos entre los estudiantes. Como resultados más significativos podemos destacar que se observó un aumento en el conocimiento de los ODS, que antes de la participación en el proyecto se reducía a aproximadamente un 50%. El cuestionario final indicó un cambio de hábitos y comportamientos que permiten pequeños ahorros de agua o electricidad por parte de los estudiantes. Los resultados indicaron también una alta motivación de los alumnos con estas actividades. Con respecto a los estudiantes preuniversitarios, el 60% indicó que le gustó la colaboración con universitarios y al 65% le gustó las actividades realizadas. Además, el 72% de los estudiantes consideraba que había aprendido cosas interesantes. Hay que indicar que, en algunas de las asignaturas, se evaluaron también los proyectos y materiales creados por los propios estudiantes universitarios, observándose, también, una alta implicación y motivación por parte de los mismos.

## 4. CONCLUSIONES

En este capítulo hemos descrito algunas de las actividades realizadas en el marco del proyecto APS22.2003 "Aprendizaje BAsado en SErvicio DE ODS relacionados con una PROducción y consumo responsableS (A-BASE-DE-PROS)". Este proyecto de aprendizaje basado en servicio tenía como objetivo dar a conocer los Objetivos de Desarrollo Sostenible entre estudiantes de educación secundaria. Para ello, estudiantes de la Universidad Politécnica de Madrid llevaron a cabo distintas actividades relacionadas, fundamentalmente, con el ODS 12 de producción y consumos responsables (aunque también con otros, como el ODS 6 agua limpia y saneamiento o el ODS 2 hambre cero), en 6 centros de educación secundaria de la Comunidad de Madrid, exponiendo los trabajos que habían realizado en la universidad, creando bioplásticos, diseñando recipientes reciclados y disoluciones para cultivos hidropónicos o trabajando como monitores de una *escape room* o de diversos talleres.

En general, hemos observado un alto grado de satisfacción entre todos los participantes, tanto en el caso de los estudiantes universitarios que llevaban a cabo el aprendizaje basado en servicio como en el de los estudiantes de secundaria que jugaban el papel de destinatarios finales del mismo.

## 5. AGRADECIMIENTOS

## 6. REFERENCIAS